\documentclass[aps, apl,preprint]{revtex4-1}
\usepackage{amsmath, amsthm}
\usepackage{amssymb}
\usepackage{tikz}
\usepackage{lipsum}
\usepackage{amsfonts}
\usepackage{setspace}
\usepackage{amssymb}
\usepackage{graphicx}
\usepackage{color}
\usepackage{array}
\usepackage{appendix}
\usepackage[justification=justified,singlelinecheck=false]{caption}
\usepackage{subfigure}
\usepackage{afterpage}
\usepackage{dcolumn}
\usepackage{ragged2e}
\usepackage{url}
\usepackage{multirow}
\usepackage{enumerate}
\usepackage{epstopdf}
\usepackage[margin=2cm]{geometry}
\begin{document}\title{Exploration of the momentum of ferroelectric domain walls via molecular dynamics simulations }
\date{\today}
\author{ Shi Liu, Ilya Grinberg, and Andrew M. Rappe}
\affiliation{The Makineni Theoretical Laboratories, Department of Chemistry,University of Pennsylvania, Philadelphia, PA, 19104-6323 }
\begin{abstract}
The motion of ferroelectric domain walls (DW) is critical for various technological applications of ferroelectric materials. One important question that is of interest both scientifically and technologically is whether the ferroelectric DW has momentum.  To address this problem, we performed canonical ensemble molecular dynamics simulations of 180$^\circ$ and 90$^\circ$ DWs under applied electric field. Examination of the evolution of the polarization and local structure of DWs reveals that they stop moving after the removal of electric field. Thus, our computational study shows that ferroelectric domain walls do not have  momentum. 
\end{abstract}
\maketitle
Ferroelectric materials have been studied intensively due to their numerous important technological applications in electronics, optics, and acoustics.~\cite{Lines77, Scott00, Scott07p954, Rabe07}    In many cases, ferroelectrics adopt a multi-domain state where domains with polarization uniformly oriented in one direction are surrounded by domains with polarization pointing in other directions. The boundary separating regions of different polarity is called a domain wall (DW).~\cite{Lines77} The DW can be moved by external electric field and stress, causing one region to grow. Therefore, controlling DW motion is critical to applications of ferroelectric materials such as non-volatile random access memory.~\cite{Yang08, Kim09p262902, Seidel09, Lu12} Though ferroelectric materials have been studied extensively for more than fifty years, the microscopic understanding of how different types of DWs form and move remains incomplete. One unanswered question  is whether ferroelectric DWs exhibit real momentum. It is generally assumed that a ferroelectric DW has real inertia, whereas the magnetic DW has no momentum.~\cite{Catalan12} However, there are no conclusive observations, either experimentally or computationally, to support the existence of momentum of ferroelectric DWs. On the other hand, recent experimental studies reveal that magnetic domain walls can exhibit significant momentum.~\cite{Thomas12} \\

To model the dynamics of DWs, the simulation of a large system at finite temperature is usually required. To enable such simulations, we have recently developed a new type of interatomic potential based on bond-valence theory.~\cite{Brown09, Grinberg02,Shin05, Shi12, Shi13} The model potentials for two ferroelectric materials, PbTiO$_3$ and BiFeO$_3$, have been parametrized  based on first-principles results.~\cite{Shi12, Shi13} The optimized potential is accurate for  both constant volume ($NVT$) and constant pressure ($NPT$) conditions and sufficiently efficient for large-scale ($\approx$1,000,000 atoms) molecular dynamics (MD) simulations. \\

In this work, we use MD simulations to study the momentum of DWs in the classic PbTiO$_3$ ferroelectric. As shown in Figure 1, the 180$^\circ$ DW (the wall separating regions with antiparallel polarization) is constructed with a $24\times8\times8$ supercell with polarization aligned along $z$ axis. The 90$^\circ$ DW (the wall separating regions with perpendicular polarization) is modeled with a $40\times40\times40$ supercell with alternating domains having polarization in $xy$ plane.
To explore the momentum of DWs, $NPT$ MD simulations with a Parrinello-Rahman barostat are performed as follows: first, the DW motion is initiated by applying an external electric field for a period of time; then the field is turned off allowing the DWs to evolve freely. If the DW has real momentum, it will keep moving after the electric field stimulus is removed. Given that accurate determination of the DW position is critical for the evaluation of momentum, we first determined the thickness of DWs. As shown in Figure 2, we calculated the averaged polarization for each layer of cells across the DWs at temperatures from 10~K to 240~K. We found that the 180$^\circ$ DW is 1-2 unit cells thick, while the width of 90$^\circ$ DW is 4-5 unit cells (the half width of the polarization profile around the domain boundary) at finite temperature. \\

 The change of the overall polarization of the supercell directly reflects the DW motion. Figure 3 presents the change of $P_z$ for 180$^\circ$ DW under various pulses at 220~K. As the electric field is applied along $+z$ direction, the magnitude of $P_z$ increases, suggesting the field-driven movement of DW along $+x$. 
Once the field is turned off, two types of equilibrations are observed: 1) the magnitude of $P_z$ is reduced until it reaches equilibrium, for example for $E=1.8$~MV/cm, $t_E$=4~ps; 2) the magnitude of $P_z$ first decreases, then increases by a small amount and eventually equilibrates, for example $E=2.2$~MV/cm, $t_E$=4~ps.  

To elucidate the origins of these two types of equilibrations, we then carried out a  close examination of the structure of 180$^\circ$ DW by analyzing the evolution of the local polarizations in the supercell. Figure 4 presents the change of the polarization profile in response to a 4~ps-long electric field pulse. The drop of $P_z$ at the instant of field removal results mainly  from the reduction in the magnitudes of local dipoles that were aligned with the applied field ($+z$) and the increase in the magnitude of the dipoles pointing toward $-z$.  When the field is removed, both types of dipoles return to their zero-field values.  More importantly, both the polarization profiles and the visualized domain patterns show that the position of the DW does not change once the field is removed, which suggests that the 180$^\circ$ DW does {\em not} have  momentum. The two types of polarization responses are actually caused by the growth and annihilation of a nucleus at the domain boundary. At 4~ps, we see  from Figure 4(a) that the local dipoles in layer $N=16$ were partially switched at the time of field removal. Similarly, in Figure 4(b), it is observed that the switching process in layer $N=16$ has already started, but has not yet finished. This means the domain wall is not flat during the motion. Figure 5 shows the changes of the dipoles in these two layers in the absence of electric field. We can see that at 4~ps, the nucleus (number of red squares) in $N=16$ ($E$=1.8 MV/cm) is small and will eventually disappear, resulting in the reduction of the polarization (Type 1 response). By contrary, the size of nucleus in layer $N=16$ ($E$ = 2.2 MV/cm) is already large; therefore, the nucleus can keep growing until the whole layer becomes $+z$ polarized. This spontaneous switching process is responsible for the increase of the polarization (Type 2 response). We therefore suggest that after the electric field is turned off, only the layers in which the size of nucleus exceeds the critical size will finish the switching process in the absence of electric field. \\ 

The evolution of $P_x$ for 90$^\circ$ DW in response to different electric-field pulses at 200~K is shown in Figure 6. Since the 90$^\circ$ DW energy is known to be about four times lower than 180$^\circ$ DW energy~\cite{Meyer02}, much smaller electric fields were applied along $+x$ direction. The increased magnitude of $x$-component of total polarization under electric field again indicates the movement of DWs driven by field. Similar to what was found for 180$^\circ$ DWs, the removal of electric field results in a decrease of  $P_x$. Figure 7 shows the changes of domain patterns and $P_x$ profiles for a 4~ps 0.4~MV/cm electric pulse. The motion of 90$^\circ$ DW (highlighted as red broken line) is evident from 0~ps to 4~ps. The overall result is the increase of the area of  field-favored domains with the sacrifice of domains with dipoles oriented oppositely to the field.
Again, the domain patterns from 4~ps to 20~ps did not change, suggesting that 90$^\circ$ DW also does {\em not} have momentum. The main effect after the field removal is the structural relaxation leading to reduction of the $x$-component polarizations. \\

In summary, we have explored the motions of both 180$^\circ$ and 90$^\circ$ domain walls in PbTiO$_3$, subjecting multi-domain samples to electric field pulse via molecular dynamics simulations. The analysis of changes of polarization and evolution of domain patterns reveal that both types of domain walls will stop moving when the electric field is turned off.  Therefore, these ferroelectric domain walls do not have  momentum. \\

\indent S.L. was supported by the NSF through Grant DMR11-24696. I.G. was supported by the US DOE under Grant DE-FG02-07ER46431. A.M.R. was supported by the US ONR under Grant N00014-12-1-1033. Computational support was provided by the US DOD through a Challenge Grant from the HPCMO, and by the US DOE through computer time at NERSC. We thank Prof. Marty Gregg for simulating discussions.\\

\newpage
\begin{figure}[!] 
   \centering
   \includegraphics[scale=0.8]{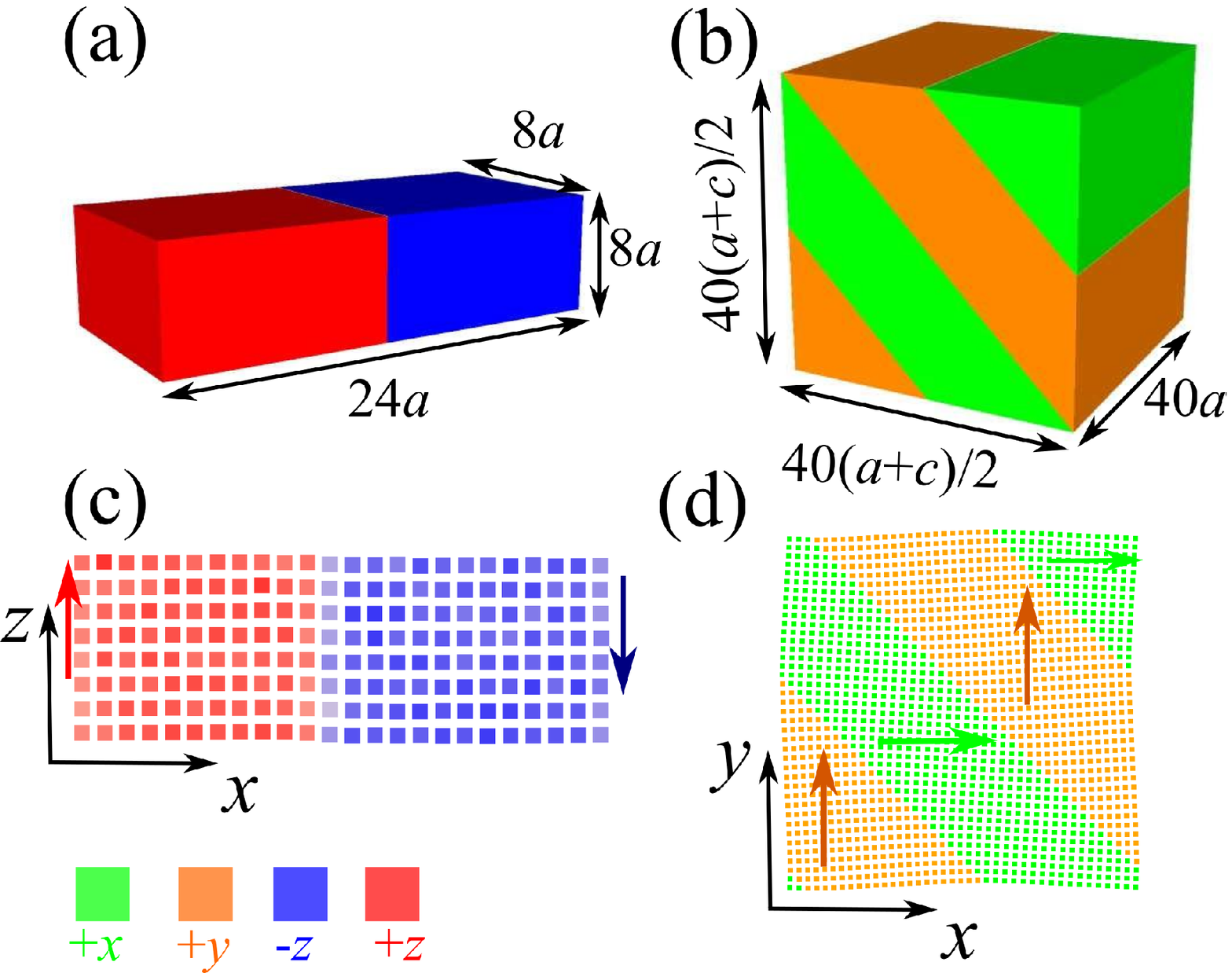} 
   \caption{(Color online) Domain walls in PbTiO$_3$. (a) $24\times8\times8$ supercell used for 180$^\circ$ DW. (b) $40\times40\times40$ supercell used for 90$^\circ$ DW. (c) domain pattern of 180$^\circ$ DW in $xz$ plane. (d) domain pattern of 90$^\circ$ DW in $xy$ pane. Each cell is colored based on the direction of the dipole: green for $+x$, orange for $+y$, red for $+z$, and blue for $-z$.}
   \label{Figure1}
   \vspace{-0.2cm}
   \end{figure}
   
 \begin{figure}[!] 
   \centering
   \includegraphics[scale=0.45]{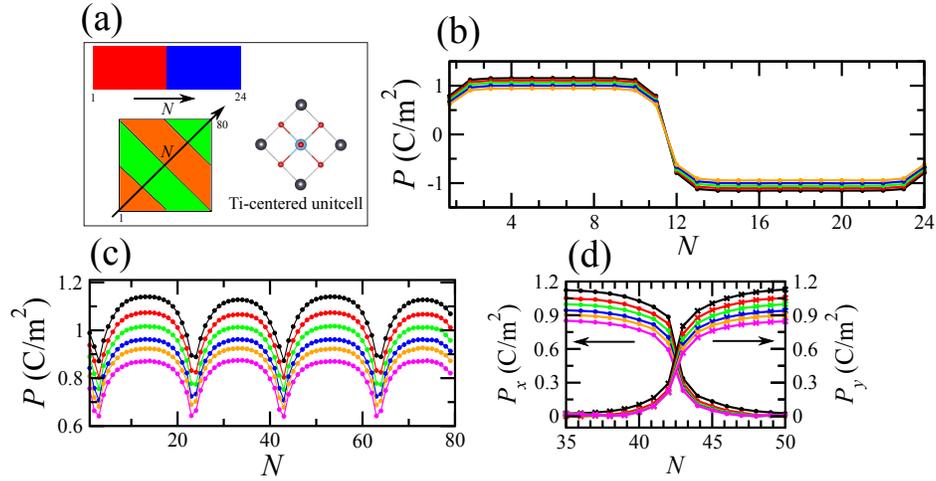} 
   \caption{(Color online) (a) Illustration of our choice of layer index, $N$, in DWs. The Ti-centered unit cell is used for local polarization calculation. (b)-(c) Temperature-dependent polarization profiles across the 180$^\circ$ DW and the 90$^\circ$ DW. (d) Temperature-dependent $x$-component and $y$-component polarization profiles across the 90$^\circ$ DW from layer $N$=35 to layer $N$=50. Lines with different color represent different temperatures: black, 10~K; red, 100~K; green, 160~K; blue, 200~K; orange, 220~K; magenta, 240~K.    }
   \label{Figure2}
   \end{figure}
 
 \begin{figure}[!] 
   \centering
   \includegraphics[scale=0.5]{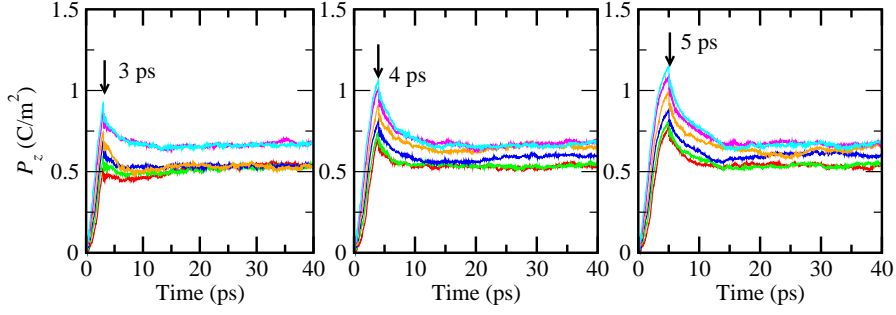} 
   \caption{(Color online) The evolution of $z$-component  polarization of 180$^\circ$ DW in response of electric pulse. Lines with different color represent different electric fields: red, 1.8~MV/cm; green, 2.0~MV/cm; blue, 2.2~MV/cm; orange, 2.4~MV/cm; magenta, 2.6~MV/cm; cyan, 2.8~MV/cm. The electric field is turned off at 3~ps, 4~ps, and 5~ps, respectively.}
   \label{Figure2}
   \end{figure}
 
 \begin{figure}[!] 
   \centering
   \includegraphics[scale=0.5]{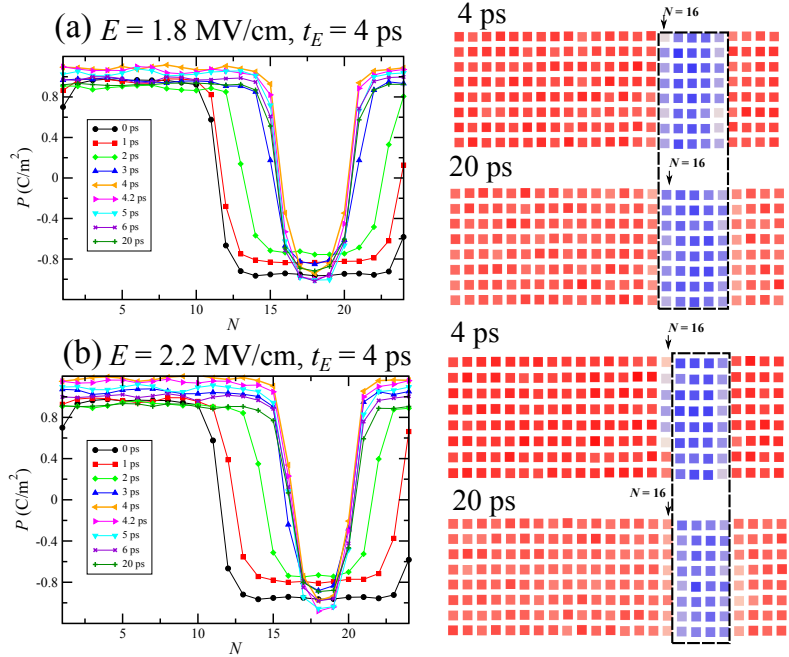} 
   \caption{(Color online) Evolution of the polarization profiles (left) and domain patterns (right) of 180$^\circ$ DW.  }
   \label{Figure4}
   \end{figure}
     
 \begin{figure*}[t] 
   \centering
   \includegraphics[scale=0.4]{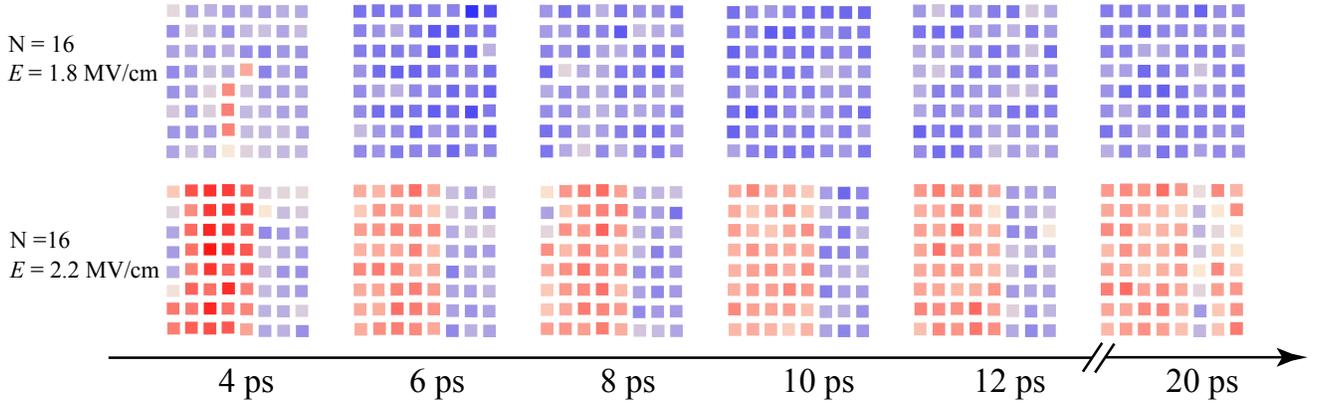} 
   \caption{(Color online) Schematic representation of nucleus annihilation (top) and growth  (bottom) at domain boundary. The electric field is turned off at 4~ps.}
   \label{Figure5}
   \end{figure*}
   
 \begin{figure}[!] 
   \centering
   \includegraphics[scale=0.45]{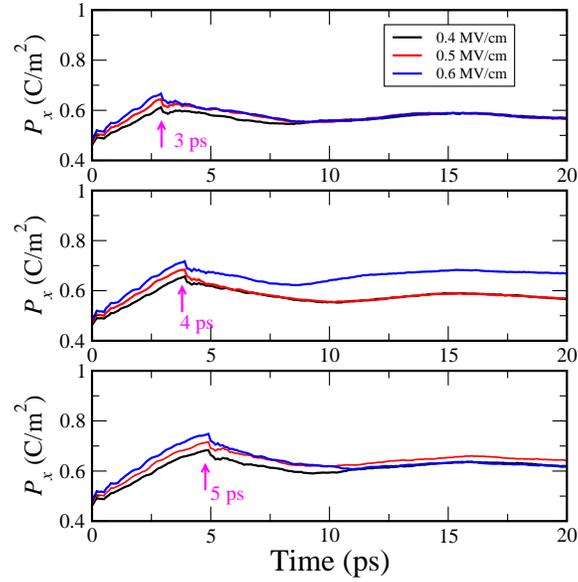} 
   \caption{(Color online) The evolution of $x$-component  polarization of 90$^\circ$ DW in response of electric pulse. The electric field is turned off at 3~ps, 4~ps, and 5~ps, respectively.}
   \label{Figure6}
   \vspace{-0.2cm}
   \end{figure}

\begin{figure}[!] 
   \centering
   \includegraphics[scale=2.00]{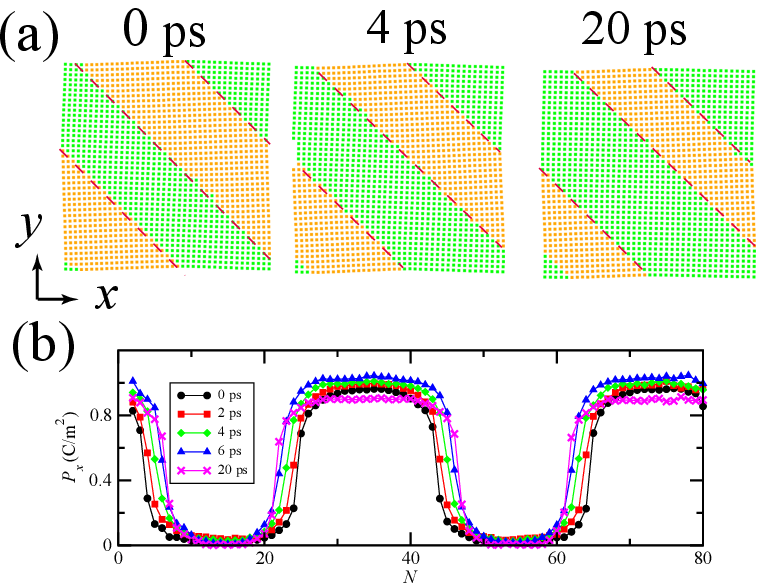} 
   \caption{(Color online) Evolution of (a) the domain patterns and (b) polarization profiles of 90$^\circ$ DW experiencing 4~ps 0.4~MV/cm electric pulse.  }
   \label{Figure6}
   \vspace{-0.2cm}
   \end{figure}

\begin{thebibliography}{99999}
\bibitem{Lines77} M. E. Lines and A. M. Glass, {\em Principles and Applications of Ferroelectrics and Related Materials} (Clarendon Press, Oxford, 1977). 
\bibitem{Scott00} J. F. Scott, {\em Ferroelectric Memories} (Springer, Berlin, 2000)
\bibitem{Scott07p954} J. F. Scott, Science {\bf 315}, 954 (2007).
\bibitem{Rabe07} K. Rabe, Ch. H. Ahn, and J.-M. Triscone, {\em Physics of Ferroelectrics:
A Modern Perspective}, Volume 105 of Topics in Applied Physics (Springer, Berlin, 2007).
\bibitem{Yang08} S. M. Yang, J. Y. Jo, D. J. Kim, H. Sung, T. W. Noh, H. N. Lee, J.-G. Yoon, and T. K. Song, Appl. Phys. Lett. {\bf92}, 252901 (2008). 
\bibitem{Kim09p262902}T. H. Kim, S. H. Baek, S. M. Yang, S. Y. Jang, D. Ortiz, T. K. Song, J.-S. Chung, C. B. Eom, T. W. Noh, and J. G. Yoon, Appl. Phys. Lett. {\bf 95}, 262902 ,(2009).
\bibitem{Seidel09} J. Seidel {\em et al}, Nature Mater. {\bf 8}, 229 (2009).
\bibitem{Lu12} H. Lu, C.-W. Bark, D. Esque de los Ojos, J. Alcala, C. B. Eom, G. Gatalan, A. Gruverman, Science {\bf 336}, 59 (2012).
\bibitem{Catalan12} G. Catalan, J. Seidel, R. Ramesh, J. F. Scott, Rev. Mod. Phys. {\bf 84}, 119 (2012). 
\bibitem{Thomas12} L. Thomas, R. Moriya, C. Rettner, S.S.P. Parkin, Science {\bf 330}, 1810 (2010). 
\bibitem{Brown09} I. D. Brown, Chem. Rev. {\bf109}, 6858 (2009).
\bibitem{Grinberg02} I. Grinberg, V.R. Cooper and A.M. Rappe, Nature {\bf 419}, 909 (2002).
\bibitem{Shin05} Y.-H. Shin, V. R. Cooper, I. Grinberg and A. M. Rappe, Phys. Rev. B {\bf 71}, 054104 (2005). 
\bibitem{Shi12} S. Liu, H. Takenaka, T. Qi, I. Grinberg, and A. M. Rappe, arXiv:1211.5166 (2012). 
\bibitem{Shi13} S. Liu, I. Grinberg, and A. M. Rappe, J. Phys. Cond. Matt. {\bf 25}, 102202 (1-6) (2013).
\bibitem{Shin07}Y.-H. Shin, I. Grinberg, I.-W. Chen and A. M. Rappe,Nature {\bf 449}, 881 (2007).
\bibitem{Meyer02} B. Meyer and D. Vanderbilt, Phys Rev B {\bf65}, 104111 (2002).
\end{thebibliography}
\end{document}